\documentclass[a4paper,11pt]{article}
\usepackage{aaskaiid,aas_macros}
\usepackage{ulem}
\usepackage{wrapfig}
\usepackage{xcolor}
\usepackage{orcidlink}
\usepackage{hyperref}
\usepackage[nameinlink,capitalise]{cleveref}
\usepackage{amsfonts,amsmath,bm,mathtools}
\usepackage{ulem,comment,siunitx}
\DeclareSIUnit\Jy{Jy}
\usepackage{tikz}
\usetikzlibrary{arrows.meta}
\usepackage{textcomp}

\def\jat#1{\textcolor{black}{#1}}

\def\cht#1{\textcolor{black}{#1}}

\def\csst#1{\textcolor{black}{#1}}

\def\dpt#1{\textcolor{black}{#1}}

\def\frt#1{\textcolor{black}{#1}}

\def\svht#1{\textcolor{black}{#1}}
\def\svhta#1{\textcolor{black}{#1}}

\def\clh#1{\textcolor{black}{#1}}
\def\todo#1{\textcolor{black}{#1}}
\def\jab#1{\textcolor{black}{#1}}
\newcommand{\clhnew}{\textcolor{black}}
\newcommand{\jabnew}{\textcolor{black}}
\newcommand{\dpnew}{\textcolor{black}}

\def\deg{$^{\circ}$}

\title{Cosmology from Clustering of Continuum Galaxies}
\ShortTitle{Cosmology from continuum clustering}

\author[1]{Jacobo\ Asorey\orcidlink{0000-0002-6211-499X}}
\author[2, 3]{Catherine L.\ Hale\orcidlink{0000-0001-6279-4772}}
\ShortName{Asorey, Hale et al.} 
\author[4]{Sebastian\ von Hausegger\orcidlink{0000-0002-6274-1424}}
\author[5]{David\ Parkinson\orcidlink{0000-0002-7464-2351}}
\author[5]{Chandra Shekhar\ Saraf\orcidlink{0000-0002-5149-4042}}
\author[6]{Jonah D.\ Wagenveld\orcidlink{0000-0003-1321-0886}}
\author[7,8,9]{Benedict\ Bahr-Kalus\orcidlink{0000-0002-4578-4019}}
\author[10,11]{Syed Faisal ur\ Rahman\orcidlink{0000-0001-9414-175X}}
\author[12,13,14]{Ziad\ Sakr\orcidlink{0000-0002-4823-3757}}

\affiliation[1]{Departamento de F\'{\i}sica Te\'orica, Centro de Astropart\'iculas y F\'isica de Altas Energ\'ias (CAPA), Universidad de Zaragoza, 50009 Zaragoza, Spain}
\emailAdd{jasorey@unizar.es}
\affiliation[2]{Institute for Astronomy, Royal Observatory Edinburgh, Blackford Hill, Edinburgh, EH9 3HJ, UK}
\emailAdd{Catherine.Hale@ed.ac.uk}
\affiliation[3]{Astrophysics, Department of Physics, University of Oxford, Keble Road, Oxford, OX1 3RH, UK}
\affiliation[4]{Rudolf Peierls Centre for Theoretical Physics, University of Oxford, Parks Road, Oxford OX1 3PU, UK}
\affiliation[5]{Korea Astronomy and Space Science Institute, 776 Daedeok-daero, Yuseong-gu, Daejeon 34055, Republic of Korea}
\affiliation[6]{Max-Planck Institut für Radioastronomie, Auf dem Hügel 69, 53121 Bonn, Germany}
\affiliation[7]{INAF, Osservatorio Astrofisico di Torino, Via Osservatorio 20, 10025 Pino Torinese, Italy}
\affiliation[8]{Dipartimento di Fisica, Universit\`a degli Studi di Torino, Via P.\ Giuria 1, 10125 Torino, Italy}
\affiliation[9]{INFN, Sezione di Torino, Via P.\ Giuria 1, 10125 Torino, Italy}
\affiliation[10]{SBASSE at Lahore University of Management Sciences, LUMS, Lahore, Pakistan}
\affiliation[11]{NCBC at NED University of Engineering and Technology, Karachi, Pakistan}
\affiliation[12]{Instituto de Física Teórica UAM-CSIC, Campus de Cantoblanco, 28049 Madrid, Spain}
\affiliation[13]{Institut de Recherche en Astrophysique et Plan\'etologie (IRAP), Universit\'e de Toulouse, CNRS, UPS, CNES, 14 Av. Edouard Belin, 31400 Toulouse, France}
\affiliation[14]{Universit\'e St Joseph; Faculty of Sciences, Beirut, BP-11514, Lebanon}

\abstract{The distribution of radio continuum galaxies is a useful, fast, and accessible probe of the matter distribution in the Universe, enlightening us about the Universe's initial conditions, the physics of dark matter, and the nature of the mysterious dark energy. However, radio continuum galaxies alone cannot easily be localised in the radial direction, and cross-identification of host sources from optical catalogues is challenging across wide area surveys. Moreover, there are several redshift-dependent properties of radio galaxy populations that all need accurate modelling to make reliable inferences about fundamental physics. These include accurate measurements of the redshift distribution of radio sources ($dN/dz$), the coupling between radio galaxies and the underlying matter distribution (quantified by the galaxy bias, $b(z)$), and the true flux distribution $N(S,z)$ of the radio sources (magnification bias). The amount of encoded cosmological information depends on the survey properties and the level of homogeneity across its footprint. In this chapter, we demonstrate the cosmological potential of a 20,000 sq. deg survey with the SKAO in AA4 configuration, using 10,000 hours of observations. Such a survey will reach $\mathcal{O}(\mu\mathrm{Jy/beam)}$ sensitivities and detect $\mathcal{O}$(300-400 million) radio sources, the largest sample of radio continuum galaxies to date. This surpasses the number of sources assumed for the previous SKA cosmology Red Book. We predict the angular clustering of such a survey, using mocks accounting for potential telescope systematics, and discuss which data corrections may be needed when these systematics cannot be accurately modelled.}

\begin{document}
\maketitle
\section{Introduction}
\label{sec:intro}

The distribution of matter in the Universe is predicted by the gravitational collapse of an initially (almost completely) smooth fluid, which through the evolution of the expanding Universe acts to generate structures on all scales. \clhnew{Commonly known as ``the cosmic web", this structure} was well established through optical spectroscopic surveys such as the 2dF Galaxy Redshift Survey \citep[2dFGRS][]{2dF2001}, demonstrating \clhnew{regions with} dense clusters, empty voids and filamentary structures. The measurement of \clhnew{the large-scale structure} through observations of galaxy positions provides an important test not just for the cosmological model ($\Lambda$CDM) as a whole, but also for the details of the physics that give rise to these structures. For example, a change in the expansion history, or a non-cold model of dark matter, or \jabnew{different} inflationary models, can all give rise to a different distribution of \clhnew{galaxies}. Moreover, \clhnew{baryonic physics} is also crucial in influencing the distribution of galaxies within the cosmic web, \clhnew{as} `ordinary' matter (which contributes the visible emission from galaxies) \dpnew{is} not solely affected by the effects of \clhnew{gravity}. Instead chemical processes, galaxy interactions and the feedback from star formation and accreting supermassive black holes are all known to be important in shaping galaxy properties and their distribution, especially on the smallest scales \citep[see e.g.][]{1999MNRAS.303..188K,Mead:2020vgs}.

\cht{Therefore, the distribution of galaxies and their large-scale structure broadly falls into two categories. On the largest scales, galaxies are biased tracers of the underlying dark matter haloes \clh{\citep[see e.g.][]{ Peebles1980, Kaiser1984, Desjacques2018}} - namely galaxies form within dark matter haloes, whose distribution is governed by the underlying cosmological parameters which determine the Universe expansion (e.g. $H_0$, $\Omega_m$, $\Omega_{\Lambda}$). On the smaller scales, the distribution of galaxies instead is governed by how galaxies form within dark matter \clhnew{haloes, where} satellite galaxies can form around a central galaxy within the halo. At these scales, baryonic physics is crucial in further understanding \clhnew{this distribution}. Traditionally, techniques such as abundance matching \clh{\citep[and sub-halo abudance matching see e.g.][]{Reddick2013, Lehmann2017}} and halo occupation distribution modelling \clh{\citep[HOD; see e.g.][]{Cooray2002, Zheng2005}} have been used to understand the relationship of galaxies forming within haloes and in understanding the connection between the properties of dark matter haloes, and the galaxies \clhnew{which inhabit them (e.g. redshifts, star formation rates etc.)}. In this work, we focus on continuum clustering \jabnew{at the} largest scales - though we will discuss the small scale effects in Section \ref{sec:hod}. }

Whilst historical measurements of the \cht{large-scale} distribution of galaxies have indicated the Universe is governed by a $\Lambda$CDM model, recent results from optical galaxy redshift surveys \jabnew{such as}
the Dark Energy Spectroscopic Instrument (DESI) \clh{indicate that (when combined with other \jabnew{probes})} the $\Lambda$CDM predictions are not a good fit to distance data \citep{2025PhRvD.112h3515A}, but these will need to be validated and confirmed with other independent \clhnew{measurements.} \clh{This requires deep surveys, over vast areas, which can probe the distribution of matter over a broad range of redshifts.}

In this vein, radio continuum galaxies can provide an important, independent sample of galaxies to trace the large-scale structure of the Universe. Radio imaging can cover large survey areas (\clhnew{to minimize cosmic variance limitations}) with fast survey speeds \citep[see e.g.][]{2020PASA...37...48M, MeerKLASSUHF}, to preferentially high redshifts and avoiding systematic effects (such as dust extinction) which can affect \clhnew{surveys} at other wavelengths. However, continuum radio surveys cannot alone provide radial information, and so the clustering needs to be measured in angular space rather than in terms of 3-dimensional Cartesian separations. This means that the projection of the cosmic web on the celestial sphere is instead relied on to recover cosmological information. Nevertheless, radio sources can still be used to provide details of the largest possible structures, both in terms of extremely wide angular separation, the evolution of the gravitational potential back in time through cross-correlation with other LSS tracers \citep[see other chapters, e.g.][]{Harrison01.2026.SKA,Spinelli02.2026.SKA} \cht{and tracing populations of galaxies across large redshifts}. 

\label{sec:radiocontpop}
\cht{Whilst the primary physical mechanisms allowing the observation of galaxies in traditional optical and near-IR surveys \clh{\citep[such as being used for several Stage III and IV cosmology surveys e.g.][]{LSST, Euclid}} are the stellar\clhnew{, gas and dust} emission of galaxies, the emission which continuum \clhnew{radio} sources trace is fundamentally different. Namely, the primary physical mechanisms responsible for radio continuum emissions are non-thermal synchrotron and thermal free–free radiation \clh{\citep[see e.g.][]{Condon1992, Hardcastle2020}}. This leads to two dominant populations: accreting supermassive black holes, known as active galactic nuclei \citep[AGN;][]{1974MNRAS.167P..31F, 1993ARA&A..31..473A, 1995PASP..107..803U, 2012MNRAS.421.1569B, Heckman2014} and star-forming galaxies \citep[SFGs; e.g.][]{Condon1992, 2003ApJ...586..794B, 2017MNRAS.466.2312D}. For bright radio AGN (known as radio loud or RLAGN), this non-thermal emission relates to high energy electrons spiraling in the magnetic fields associated with jets which are expelled from the central supermassive black hole. The picture for radio quiet AGN (RQAGN), is instead more uncertain as to the contribution of \clhnew{jets, winds or} star formation activity to the radio emission \clh{\citep[see e.g.][]{2017A&A...602A...3D, Panessa2019, White2025}}. For SFGs the non-thermal synchrotron emission is instead associated with high energy electrons spiraling in magnetic fields associated with supernova \clhnew{and} \clh{the thermal free–free radiation instead traces HII} regions.  }


\cht{At the brightest flux densities ($\geq 1$mJy at 1.4 GHz), the radio sky is dominated by RLAGN. These are typically hosted by massive ellipticals residing in group or cluster environments and are associated with dark matter halos of mass $M_{\text{halo}} \geq 10^{13} M_{\odot}$ \citep[e.g.][]{2004MNRAS.350.1485M, Magliocchetti2022, Hamlett2026}. The feedback from AGN is known to be a crucial aspect in the growth and evolution of galaxies, with feedback \clhnew{crucial to reconcile observations with predictions of the most massive galaxies from cosmologial simulations} \clh{\citep[see e.g.][]{Benson2003, Dubois2016}}. Whilst historically, such radio loud populations were categorised based on \clhnew{morphology} \citep[known as Fanaroff-Riley Type I and II AGN][]{1974MNRAS.167P..31F}, more recent works have classified radio loud AGN based on signatures which are indicative of different accretion modes onto the AGN, known as high/low excitation radio galaxies \citep[H/LERGs; see e.g.][]{Heckman2014, Hardcastle2020, Whittam2022}. At sub-mJy and $\mu$Jy levels, SFGs and radio-quiet (RQ) AGN instead dominate. \clhnew{For RQAGN, recent works \citep{Radcliffe2021, Njeri2025, White2025} are beginning to provide large enough samples to gain further understanding of the origin of their emission.} \clh{In SFGs, the non-thermal emission is commonly used as a delayed tracer of \clhnew{their star-formation rate}, unobscured by dust. This allows radio SFGs to play a pivotal role in \clhnew{reconstructing the} cosmic star formation history of the Universe in an unbiased manner \citep{Smith2021, Matthews2021b, Cochrane2023}. }}

\cht{Combined, this lack of dust obscuration and the powerful emission of AGN means that radio sources can be observed to large distances in the Universe, with recent radio surveys peaking in redshift at $z \sim1$ \citep[e.g.][]{2024MNRAS.527.3231W}. This means that \clhnew{radio surveys are a powerful tool to probe} the distribution of matter over large periods of cosmic time. We note though, that \clhnew{flux limited surveys will contain different populations which are contributing to the survey due to the differences in both redshift and intrinsic luminosity distribution of sources \citep[see e.g.][]{Whittam2022, Best2023}, as these populations also have different bias contributions, this will affect contributions to the large-scale structure \citep[see e.g.][ and Section \ref{sec:radiobias}]{Hale2018, Hale:2025wxp}.}} 


Given the potential of the \clhnew{SKA Observatory (SKAO)} to produce large area surveys, it is important to understand the capabilities of the SKAO in contributing to cosmological studies through studying the clustering of radio galaxies which are detected from large area continuum surveys. With the SKAO planned to be constructed in stages, \clhnew{currently planned to build towards the so called AA4 configuration} \citep{braun2019anticipatedperformancesquarekilometre}, in this work we consider how the extended AA4 configuration can provide a \clhnew{a survey of galaxies to measure their auto-correlation} (and by extension cross-correlation with other tracers) for testing and constraining of the cosmological model. Whilst much of the science possible has been outlined in the previous SKA chapters \citep{Jarvis2015, RedBook}, updates in our understanding of the faint radio source populations have advanced significantly with the advent of deep surveys from precursor facilities such as the Australian SKA Pathfinder \citep[ASKAP;][]{Johnston2008}, MeerKAT \citep{Jonas2009} and the LOw Frequency ARray \citep[LOFAR][]{vanHaarlem2013}. These have demonstrated that the source count models adopted by the SKA Simulated Skies ($S^3$) catalogues which were used to make cosmological predictions significantly underestimate the faint radio source populations \citep[see e.g][]{Mandal2021, Hale2023} and we have gained increasing understanding of the galaxy bias of different sub-populations of radio sources \citep[see e.g.][]{Hale2018, Mazumder2022, Hale:2025wxp}. \clhnew{These further need to be updated for the updated specifications of the SKAO and to account for its new expectations}. 

\cht{In this chapter we therefore aim to \clh{update} the science case for continuum clustering with the SKAO for cosmological studies, \clh{given the advances in the last decade,} and produce improved predictions for the expectations of a large-area survey with the SKAO. The chapter is outlined as follows: in Section \ref{sec:clusteringstats} we address the \clhnew{statistics used} to measure the clustering of galaxies \clhnew{for} cosmological analysis. \clh{In} Section \ref{sec:pathfinders} we outline the current status of wide area radio continuum surveys used for cosmological analysis \clh{and the challenges associated with cosmology from radio continuum surveys in Section \ref{sec:challenges}.} We then discuss the simulations adopted to study expectations for a 20, 000 sq deg survey with AA4 using 10,000h of observations in \clh{Section \ref{sec:sims}.} Finally, we discuss the clustering analysis we anticipate obtaining with the SKAO in Section \ref{sec:properties}, before summarising in Section \ref{sec:summary}. }


\section{Angular Clustering Statistics for Galaxy Surveys}
\label{sec:clusteringstats}
\cht{As discussed in Section \ref{sec:intro}, whilst spectroscopy can directly trace the position of galaxies \clhnew{within the cosmic web}, radio continuum studies \clhnew{alone are restricted to} the statistical distribution of the projected density field \clhnew{for galaxy clustering studies}. Such \clhnew{studies} can both be \clhnew{measured} in real space using the distribution of galaxies as a function of projected angular separations \clhnew{or, as done in many cosmological analysis, in Fourier space}. For both studies, key is \clhnew{the} measurement of the projected density contrast (\clh{$\delta(\hat{n})$}, i.e. the excess density of sources), given by:}
\begin{equation}
    \delta(\hat{n})=\frac{N_g(\hat{n})}{\overline{N}_g}-1
    \label{eq:density}
\end{equation}
where $N_g(\hat{n})$ is the number of galaxies in pixel $\hat{n}$ and $\overline{N}_g$ is the average number of galaxies per pixel. If the number counts per pixel is incomplete due to systematics, we redefine \clhnew{a weighted galaxy number distribution} $N_g(\hat{n})=N_g(\hat{n})/w_g(\hat{n})$ and $\overline{N}_g=\langle N_g(\hat{n})/w_g(\hat{n})\rangle$, with  the map of incompleteness weights, $w_g(\hat{n})$.

In real space, clustering studies trace this density contrast through measurement of the angular {two-point angular correlation function,(\clhnew{$\omega(\theta)$}, which quantifies the excess probability to observe galaxies at a given angular separation, \( \theta \), compared to randomly distributed sources. A widely used method for calculating (\clhnew{$\omega(\theta)$}) from a catalogue \clhnew{of galaxies} is the estimator from \citet{LandySzalay1993}:
\begin{equation}
    \omega(\theta) = \frac{\overline{DD}(\theta) - 2 \overline{DR}(\theta) + \overline{RR}(\theta)}{\overline{RR}(\theta)},
    \label{eq:LS}
\end{equation}
In this equation, \( \overline{DD}(\theta) \) represents the \clhnew{normalised} pair count \clh{of observed galaxies} within the data catalogue at a particular angular scale \( \theta \), \( \overline{RR}(\theta) \) is the \clhnew{normalised} pair count in an associated random catalogue, and \( \overline{DR}(\theta) \) denotes the \clhnew{normalised} pair count between the data and random catalogues. The random source catalogue is designed to reflect variations in object detectability across the survey while exhibiting no intrinsic clustering \clh{\citep[see e.g.][]{Hale2024}}. } \jat{These random catalogues can also be used to build weighted angular masks, $w_g(\hat{n})$, that can be used to correct the definition of density contrast in Equation \ref{eq:density} when observing systematics may introduce non cosmological clustering signal.} 

In Fourier space, the density contrast is decomposed in spherical harmonics \citep[e.g.][]{Peebles1973}:
\begin{equation}
    \delta(\hat{n})=\sum_{\ell}\sum_{m=-\ell}^{\ell}{a_{\ell m}Y_{\ell m}({\hat{n}})}
\end{equation}
from which we can define the angular power spectrum, $C_\ell$, given by the coefficients of the expansion:
\begin{equation}
\left<a_{\ell m}a_{\ell' m'}^*\right>=\delta_{\ell \ell'}\delta_{m m'}C_{\ell}.
\label{eq:cl_def}
\end{equation} 

The angular power spectrum  provides an alternative way to describe the clustering of objects on the sky. It is the Fourier transform of \clhnew{$\omega(\theta)$} and represents the amplitude of the clustering signal on different angular scales. While \clhnew{$\omega(\theta)$} is more intuitive \clh{in terms of the scales considered}, $C_\ell$ can be more directly computed from theoretical models, and the multipoles $\ell$ are (almost) statistically independent for wide surveys. For \clh{a review} on the modelling of angular clustering for surveys with photometric redshifts, see \cite{2011MNRAS.414..329C}.

\clhnew{Whilst cosmological models} cannot predict the individual positions of galaxies, \clh{they can instead be used to model} their statistical clustering properties. The density contrast at a given angular position $\hat{n}$, $\delta(\hat{n})$, is \clh{essentially} the projection of the 3D field of inhomogeneities within the kernel given by the redshift distributions of our radio galaxy population. This relation is given by:
\begin{equation}
    \delta(\hat{n})=\int{n(z)\delta(z,\hat{n})dz}
    \label{eq:delta_th}
\end{equation}
where $n(z)$ is the normalized redshift distribution of our sample. Therefore, by projecting the large-scale structure information we lose part of the cosmological input. However, if we were able to do a tomographic analysis we could recover the 3D information embedded in our galaxy population \citep{2012MNRAS.427.1891A,Camera2018}. 

Combining Equations \ref{eq:cl_def} and \ref{eq:delta_th} we derive a theoretical expression for the angular auto-correlations:
\begin{equation}
    C_\ell=\frac{2}{\pi}\int{dkk^2P(k)\psi_\ell^2(k,z)}
    \label{eq:cl_theory}
\end{equation}
where $P(k)$ is the 3D linear matter power spectrum and:
\begin{equation}
    \psi_\ell=\int{dz\,n(z)b(z)D(z)j_\ell(k\chi(z))}
    \label{eq:win_gg}
\end{equation}
is the window function for galaxy-galaxy clustering. It depends on the redshift distribution $n(z)$ along which we are projecting the 3D clustering, the population bias $b(z)$ (see Section \ref{sec:radiobias}), the linear growth factor $D(z)$ of matter, which is cosmology dependent, and the spherical Bessel function of the first kind $j_\ell(k\chi(z))$ as the kernel that mixes scales, $k$ and radial distances, $\chi(z)$, \jabnew{for each multipole $\ell$}. Therefore, to estimate this window function, we need the redshift distribution of the galaxy sample, $n(z)$, and the galaxy bias, $b(z)$, \jat{that relates the observable source overdensity to the underlying matter overdensity.}

The measurements of these power spectra can be \clhnew{further} affected by gravitational lensing which can induce magnification bias. Magnification bias, depending on the observed data, can significantly affect clustering measurements, especially at high redshifts and large angular scales \citep{LoVerde2008}. In a radio continuum survey, the effect of gravitational lensing can modify the observed flux and the solid angle element of the observed sky patch. The flux increase can lift otherwise undetectable sources above the survey minimum flux limit, while the stretching of the sky can alter the number counts of sources per unit. The net effect on the observed overdensity can be written as:
\begin{equation}
\delta_{obs} = \delta + \delta_\mu,
\end{equation}
where the magnification biased-induced term is given by:
\begin{equation}
\delta_\mu = (\alpha - 1)\kappa,
\end{equation}
with \( \kappa \) being the lensing convergence and \( \alpha \) the logarithmic slope of the integral source counts:
$N(>S) = C S^{-\alpha}$.
Thus, the total observed correlation includes the actual clustering and magnification-bias-induced modification. The observed angular power spectrum is expressed as:
\[
C_\ell^{\mathrm{obs}} = C_\ell^{gg} + 2\,C_\ell^{g\mu} + C_\ell^{\mu\mu},
\]
where $C_\ell^{gg}$ is the intrinsic term of galaxy clustering given by Eq. \ref{eq:cl_theory}, $C_\ell^{g\mu}$ is the galaxy-magnification cross-power spectrum and $C_\ell^{\mu\mu}$ is the magnification auto-power spectrum.  The effect can become especially relevant for \jat{deep} wide-area surveys \citep{LoVerde2008, 2015ApJ...802...64B}.  Observational cross-correlations with CMB lensing and sub-mm catalogues demonstrate that magnification can produce non-negligible enhancements in measured amplitudes, so forecasts and bias estimates for radio continuum surveys must include this term to avoid biased cosmological inferences \citep{LoVerde2008,2015ApJ...802...64B,Allison2015}. However, in this work we focus on our ability to extract cosmological information from the clustering of continuum galaxy samples besides the surveying systematics so we do not include this effect in our analysis.

{{Assuming linear theory and Gaussianity on the intrinsic probability of the density field, we define the covariance matrix of the auto-correlation angular power spectra in spherical-harmonic \clh{space}:}}
\begin{equation}
    \text{Cov}(C_\ell, C_{\ell^{'}})=\frac{2}{(2\ell+1)\Delta\ell f_{\text{sky}}}(C_\ell+N_\ell)^2\,\delta_{\ell\ell^{'}}
    \label{eq:cov}
\end{equation}
where $\Delta\ell$ is multipole binwidth, $f_{\text{sky}}$ is the fraction of sky given by the survey footprint and $N_\ell$ the shot noise term given by the inverse of the number of galaxies per steradian.


\subsection{\cht{Galaxy bias in Radio Continuum surveys}}
\label{sec:radiobias}

\clh{As outlined in Equations \ref{eq:cl_def} and \ref{eq:win_gg}, key to modelling the bias of radio sources, is understanding of the bias \clhnew{evolution} within a given radio survey.}
\frt{However, as continuum survey catalogues will be \clhnew{a mixture of different source populations (varying dependent on the survey limits applied)}, the appropriate parameter to use is an effective bias that averages the halo bias over the halo occupation of the population with different source types \citep{Sheth2001,Blake2004, Wilman2008, Rahman2015}.}
\begin{equation}
b_{\rm eff}(z) \;=\; \frac{\int {\rm d}M \; b_h(M,z)\; N(M,z)}{\int {\rm d}M \; N(M,z)}.
\end{equation}
\frt{Here $b_h(M,z)$ is the halo bias \clh{for a given halo mass and redshift} and $N(M,z)$ represents the halo occupation or number density of sources as a function of halo mass and redshift. In practice, continuum samples contain the multiple classes (AGN, SFGs) \cht{as described in Section \ref{sec:radiocontpop}}. A convenient and often used phenomenological form is to write the effective bias as a sum over populations:}
\begin{equation}
b_{\rm eff}(z) \;=\; \sum_i f_i(z)\, b_i(z),
\end{equation}
\frt{where $f_i(z)$ is the redshift-dependent fractional contribution of population $i$ and $b_i(z)$ is its characteristic bias. This population decomposition underpins multi-tracer strategies and forecasts for radio surveys \citep{Raccanelli2014,Maartens2015,Bernal2019}.}


\jab{Wide-area continuum surveys are well suited to constrain large-scale modes because they provide very large survey volumes. Multi-tracer strategies, such as cross-correlations between continuum samples and CMB maps, targeting the Integrated Sachs-Wolfe signal (ISW), help isolate large-scale signals, see \citet{Rassat2007, Rahman2015, 2022MNRAS.517.3785B,2024A&A...681A.105N}. These strategies also allow us to place further constraints on primordial non-Gaussianities of the local type, $f_{NL}$, through the dependency of the scale dependent galaxy bias $\Delta b(k,z)$ \citep{Raccanelli2014}. Cross-correlating the information from tomographic redshift bins in high redshift samples enhances the constraints on $f_{\text{NL}}$ as well as using tracers from different surveys \citep{2015PhRvD..92f3525A}.}

\section{Radio Cosmology from Pathfinder and Precursor Facilities}
\label{sec:pathfinders}


The key requirements for any large-scale structure survey are for the largest possible area (preferably contiguous with few holes or gaps) and the highest possible number density, ideally with \clhnew{redshift information}. As outlined in Section \ref{sec:intro}, radio continuum surveys cannot alone provide redshifts and instead we are benefitted that \clhnew{large radio continuum surveys} can be conducted quickly, \clhnew{with a} reasonable field-of-view ($> \mathcal{O}(1)$ sq. deg). The only limitations are the observation time required to meet the target area \clhnew{and sensitivity}, and \cht{sufficient angular resolution to overcome} the confusion limit \citep{2012ApJ...758...23C,2021MNRAS.506.4121A}. \clhnew{Multi-frequency information are often not such stringent requirements, as the energy spectrum of the sources provide little extra information when sampling cosmological density fields}. \clhnew{Whilst source properties such as the morphology \citep[see e.g.][]{Morabito2025} can aid in the classification of radio sources into sub-populations (e.g. AGN and SFGs) provided the resolution is high enough, \cht{redshift information and source property information (useful for multi-tracer analysis) is typically limited to regions with \clhnew{ample} multi-wavelength data. Therefore, whilst redshifts can be useful for tomographic cross-correlations}, the utility of these from a pure continuum survey is limited.}
Here we summarise \clhnew{some of the large-area radio continuum surveys that are suitable for cosmological analysis} and have been or are planned to be \cht{conducted by the commencement of operations of the SKAO, alongside existing angular clustering based analysis}.
\dpt{\begin{itemize}
\item NVSS - the \textbf{NRAO VLA Sky Survey} \citep{condon1998nrao}, is a 1.4 GHz survey on the VLA of the sky north of $\delta >-40$\deg, detecting 1.8 million sources between 1993 and 1997. \cht{Angular clustering of NVSS sources have been well studied \citep[e.g.][]{Blake2004}, and demonstrated challenges with multi-component source clustering at the smallest angular scales}.
\item \cht{LoTSS} - the \textbf{LOFAR Two-metre Sky Survey} \citep{2017A&A...598A.104S} is an ongoing survey between 128-160 MHz using LOFAR which has produced the largest catalogue of radio sources from a survey to date, \clhnew{$\mathcal{O}$(14 million) in LoTSS-DR3 \citep{Shimwell2026}}. \clhnew{Covering, $\sim$88 \% of the northern sky, existing cosmological clustering analysis has been performed over wide areas with previous data releases \citep[e.g.][]{Siewert2020, Hale2024}}. \cht{Moreover, deeper imaging over smaller fields have also been observed \citep[LoTSS Deep, e.g.][]{Tasse2021, Shimwell2025} which have also been used to study the clustering of different sub-populations \citep[see e.g.][]{Hale:2025wxp} utilising source classification from multiwavelength data \citep[see][]{Best2023}.}
\item VLASS - the \textbf{Very Large Array Sky Survey} \citep{2020PASP..132c5001L} is a 2-4 GHz survey covering the sky north of $\delta >-40$\deg (the same as NVSS), but relatively deeper and at higher frequency. \clh{Observing} from 2017 to 2025, and is expected to catalogue $\sim$10 million sources.
\item GLEAM-X - the \textbf{GaLactic and Extragalactic All-sky Murchison Widefield Array eXtended} survey \citep{2022PASA...39...35H, Ross2024} is a survey at 72–231 MHz, covering the sky south of declination $\delta < +40$\deg. It is expected to detect $\sim$1.7 million sources and the angular clustering of sources have been studied in \cite{Venville2024}.
\item \cht{RACS - the \textbf{Rapid ASKAP Continuum Survey} \citep{2020PASA...37...48M} is \clhnew{a multi frequency survey} (888, 1400 and 1720 MHz) using short ($\sim$10min) observations with ASKAP to cover the sky south of $\delta \lesssim 40$\deg). The detected $\mathcal{O}$(2 million) sources have been studied through auto-correlation and cross-correlation with the CMB \citep{2022MNRAS.517.3785B}.}
\item EMU - the \textbf{Evolutionary Map of the Universe} survey \citep{2011PASA...28..215N,2025PASA...42...71H} is a survey centred at around 900 MHz, covering the southern sky ($\delta \lesssim 0$\deg). The survey started in May 2023, and is expected to detect around 20 million sources. \jat{The first cosmological results have been obtained using EMU pilot survey data \citep{2025PASA...42...62T}. Some early attempts to estimate the redshift distribution of EMU radio population have been done by cross-correlating EMU galaxies with the Euclid Q1 sample \citep{2025arXiv251122732P}.} 
\item Moreover, surveys such as the MeerKAT Absorption Line survey \citep[see e.g., MALS;][]{Gupta2016, Deka2024, Wagenveld2024} and the MeerKAT Large Area Synoptic Survey \citep[MeerKLASS; ][]{MeerKLASSUHF} additionally provide large area radio continuum data, despite radio continuum cosmological studies not being their primary goal. The MeerKLASS UHF Continuum Survey also demonstrates the novel on-the-fly \citep[OTF;][]{Chatterjee01.2026.SKA} using MeerKLASS UHF band observations (544-1088 MHz) with a central frequency of 816 MHz.
\end{itemize}}

\begin{figure}
    \centering
    \includegraphics[height=5.1cm]{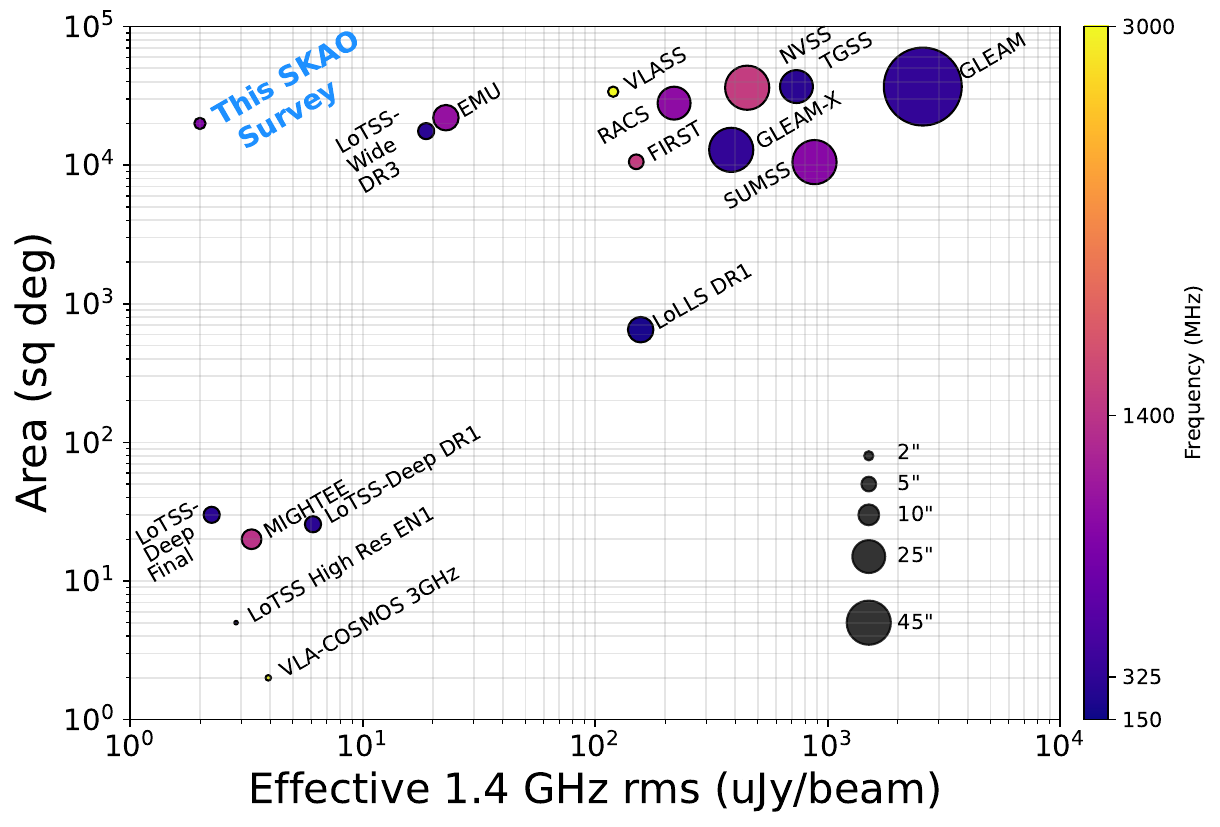}
    \includegraphics[height=5.1cm]{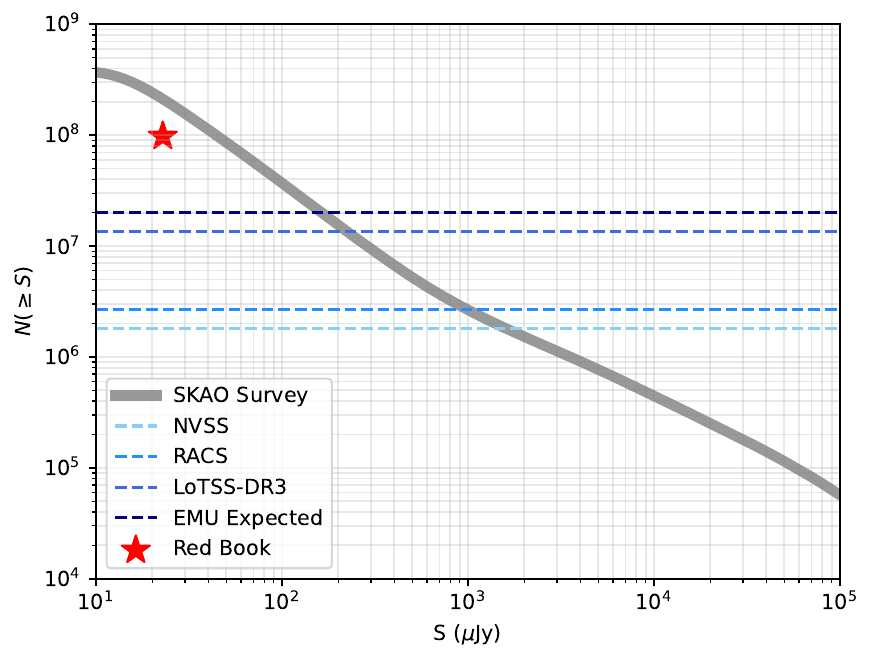}
    \caption{\cht{\textit{Left:} Comparisons of this simulated survey (blue) compared to continuum surveys from SKA precursors of the effective 1.4 GHz rms (assuming $S_{\nu} \propto \nu^{-0.7}$) and area covered. Points are coloured by their frequency and the marker size indicates the resolution. \textit{Right:} Expected source numbers with flux density limit in this survey \clhnew{(grey, using simulations outlined in Section \ref{sec:sims})} to previous/current surveys (dashed lines) and expectations from the Red Book.} }
    \label{fig:surveyscomparison}
\end{figure}



A comparison of the area/sensitivities of a number of radio surveys is presented in Figure \ref{fig:surveyscomparison}. Given these surveys, the predicted number of unique radio sources is expected to be $\sim$30-35 million before the SKAO starts operations. Therefore, for the SKAO to add value it would need to significantly improve on this number. \cht{We address predictions for such a potential SKAO survey in Section \ref{sec:sims}}.

\section{\cht{Challenges in using Radio sources to trace the LSS}}
\label{sec:challenges}
\cht{Whilst radio sources can be excellent tracers of the large-scale structure, \clhnew{offering a unique view to other wavelengths unobscured by dust, there are challenges associated with their use for large-scale structure studies which radio astronomers need to be considerate of in order to maximise the use of the SKAO for cosmology. We discuss a number of these in the following sections.}}

\subsection{Image noise and observational systematics}
\cht{One of the more important aspects for cosmology studies is to ensure the structure you are measuring is a true representation of large-scale structure and not just a manifestation of observational systematics within the survey. In large area radio surveys, the sensitivity can vary significantly across the survey \citep[see e.g.][]{Hale2021, Shimwell2022} and can be dependent on factors such as the elevation of an observation (which can be challenging for non-steerable telescopes e.g. LOFAR and SKA-Low), the ionospheric conditions (especially at the frequencies of SKA-Low) and proximity to both bright sources \citep[e.g. A-team sources, see list in][]{2020PASA...37...18W} and the \clhnew{emission from the Milky Way. Combined, these effects} can introduce apparent large-scale structure in the catalogues extracted over such areas. Moreover, the noise in radio surveys is correlated and is convolved with the PSF of the observations in the image. As such, positive noise spikes can be \clhnew{mistaken} for faint radio sources or lead to the boosting/depletion of the flux from faint sources. This will impact source detection and characterisation and thus affect relative SNR and flux density cuts across the survey. \clh{However, these variations can be accounted for with weight maps or masking \citep[e.g.][]{Hale2024}.} One advantage of radio imaging, though, is the lack of contamination in radio images from stars, dust and cosmic rays which can require significant masking and modelling to account for extinction. In radio images, stars are significantly less numerous \citep[see e.g.][]{Callingham2023} \clhnew{and instead} residual PSF artefacts around bright radio sources \clhnew{are} a more significant issue.  }

\subsection{Source Finding}
\cht{As discussed, source detection and characterisation in radio images can be affected by noise variations which has implications for large-scale structure studies. This is also impacted by the source finders themselves. Radio continuum sources finders \clhnew{typically} rely on identifying islands of emission above a background \clhnew{level} which is used to quantify the flux emitted within the island. This emission can be \clhnew{modelled} using a combination of Gaussian components \citep[see e.g. ][]{PyBDSF, Selavy, Aegean}, given radio PSFs are assumed to be Gaussian. However other source finders do not adopt such a requirement on morphology \citep[see e.g.][]{Blobcat, Profound}. More recently, machine learning has also been employed in a number of studies to extract sources \citep[e.g.][]{Vafaei2019, Stewart2024}.} \cht{The comparison of source finders has been studied in numerous works \citep[e.g.][]{Hopkins2015, 2021MNRAS.500.3821B, Boyce2023}, with \clhnew{each showing} differences in \clhnew{the} detection completeness and the accuracy in quantifying source properties. Moreover, there are often notable differences in the ability to associate all the emission from a single galaxy, when the radio emission is extended. For example, large jetted AGN can be separated into their lobes and core and be considered individual `sources'. This can lead to challenges for large-scale structure studies, causing an apparent excess of clustering at small angular scales \citep[see e.g.][]{Blake2002, Wilman2003, Hale2024}. \clhnew{These} differences between source finders \clhnew{in how they} associate sources into individual objects can affect measurements of the power spectra \citep[see e.g.][]{2025PASA...42...62T}. } \cht{As such, source association is important, else \clhnew{cosmological studies should focus on those scales without the effect of multi-component sources}. Source association can be done through visual identification \citep[see e.g.][]{Banfield2015, Williams2019, EMUzoo2025}, statistical methods \citep[see e.g.][]{Sutherland1992, McAlpine2012}, machine learning techniques \citep[see e.g.][]{Galvin2019, Alegre2024}, or a combination of methods \citep[see e.g.][]{Kondapally2021}. These, however, can require multiwavelength data to match to individual host galaxies, which is not necessarily uniform in coverage or sensitivity when such large area surveys are considered.}

\subsection{Source Characterisation and Redshift Distributions}
\cht{{Whilst the focus in this chapter is on the largest physical scales and the clustering tracing the larger scale distribution of dark matter haloes (i.e. the 2-halo clustering), \clhnew{significant} understanding can be gained from studying the clustering of different radio populations \clhnew{separately} (e.g. SFG vs AGN), at different redshifts and \clhnew{when the clustering from scales within single dark matter haloes (1-halo clustering) are studied}. Such studies require host galaxies (associated as outlined above) as well as quantification of source properties. \clhnew{In general, radio continuum imaging alone does not have sufficient information to characterize sources based on source type. This is with the exception of very high resolution imaging, where brightness temperature can be used as a proxy \cite[see e.g.][]{Morabito2025}}, \clhnew{although observable physical characteristics such as resolved jets can also help}. Therefore, deep, multiwavelength imaging is required to provide source characterisation \clhnew{and source properties, such as redshifts}. This is typically limited over the large areas considered for surveys in this work and instead these studies have typically relied on smaller imaging over well-studied fields \citep[see e.g][]{Magliocchetti2017, Hale2018, Mazumder2022, Hale:2025wxp}. }}

\cht{However, larger radio surveys have produced associated catalogues with redshifts and source properties, where available. \csst{For example, EMU Pilot data radio sources were cross-matched with Dark Energy Survey \citep[DES;][]{2021ApJS..255...20A}, Legacy Imaging Survey \citep{2019ApJS..242....8Z} and WISE $\times$ SuperCosmos \citep{2016ApJS..225....5B} photometric catalogues, obtaining photometric redshifts for $\sim 36\%$ of sources \citep{2024PASA...41...27G}. LoTSS on the other hand, cross-matched the DR2 large area survey to the DESI Legacy Imaging Surveys \citep{Dey2019} using a combination of visual inspection and statistical techniques \citep[see][]{Hardcastle2023}. This resulted in 85\% of sources having a host identified, reducing to $\sim$ 58\% of sources \clhnew{which have} an associated redshift. Efforts have further associated a fraction of the LoTSS sources to spectra from SDSS \citep[for $\sim$150 000 sources; see][]{Drake2024}. These fractions contrast $\gtrsim $90\% sources with hosts and redshift in deeper fields \citep[see e.g.][]{Kondapally2021}.}} \cht{\csst{Predicted redshift distributions from radio continuum simulations such as the Tiered Radio Extragalactic Continuum Simulation (\texttt{T-RECS}; \citealt{2019MNRAS.482....2B}) and the SKA Design Study Simulated Skies \citep[SKADS][]{Wilman2008} also have been used for cosmological inferences \citep[e.g.][]{2019MNRAS.488.5420H, 2022MNRAS.517.3785B, 2023A&A...671A..42P, Venville2024, 2025PASA...42...62T}}, but these have limitations (see Section \ref{sec:intro}).}

\cht{Moreover, the multi-tracer techniques outlined in Section \ref{sec:radiobias} require the ability to split sources into such sub-populations which will require advances in machine learning techniques to classify sources with the limited data available. Therefore, we caveat that such studies rely on assumptions to be able to make such characterisations. In this work we limit the statements made on separating source populations and restrict modelling the clustering on scales where the 2-halo term dominates. \clhnew{We note though} that with \clhnew{potential deep extragalactic} continuum surveys planned with the SKAO (see e.g. the chapter by \cite{Prandoni01.2026.SKA}) such studies will be able to be significantly advanced (see Section \ref{sec:hod}).   } 

\section{A Simulated SKAO Continuum Survey}
\label{sec:sims}

For the purposes of this chapter and evaluating the potential contribution of the SKAO to the study the large-scale structure we construct simulations of the proposed $10\,000\,{\rm h}$ survey, covering $20\,000\,{\rm deg}^2$ from the SKA Cosmology Red Book \citep{RedBook}. \clhnew{Whilst key science projects (KSPs) have yet to be defined, we use this as a reference to allow} us to implement realistic SKAO predictions with updated radio simulations to asses if the expectations of such a survey have updated since the Red Book. \cht{Therefore, we assume Band 1 of SKA-Mid in AA4 (centred at 780 MHz).} We note, though, that the Red Book proposes a commensal continuum and on-the-fly (OTF) survey. However, as we are unable to test the capabilities of OTF mode, we adopt the SKAO continuum mode predictions from the current SKA sensitivity calculator. Works from MeerKLASS have suggested that continuum imaging from such data would be possible \citep[see][]{MeerKLASSUHF}, \clhnew{however regardless as KSPs have not been defined, a wide area continuum survey could also be observed}. However, OTF imaging will lead to an elongated beam compared to longer continuum observations and the systematics in these observations may be more challenging to understand. \cht{We proceed under the assumption that (a) the imaging can be convolved to a common resolution that is not materially different to the resolutions from the SKAO calculator, (b) any systematics behave in similar ways to traditional continuum surveys or (c) a separate wide area continuum survey would be possible regardless of the availability of OTF imaging.}  
We outline the steps to construct our simulated observations in the next sections.

\subsection{Sensitivity map generation}
\label{sec:rmsmap}
We adopt a hexagonal tiling strategy to cover the Southern hemisphere ($\delta \lesssim 0$\deg), assuming a FWHM of the primary beam of of 1.22$\lambda/D$, with $D$ the diameter of an individual dish, taken as 15m for SKA-Mid. This resulted in a total of $\sim15\, 000$ pointings\footnote{We additionally produce pointings which do not observe the Milky Way, which will be considered in later Sections}. For each pointing, we utilised the SKAO continuum sensitivity calculator to obtain properties of such a pointing (rms and angular resolution), assuming a Briggs' weighting of~0 \citep[as in][]{Hale2025MIGHTEE} and {elevation\footnote{{We do not adopt the optimal elevation for each case as this will be likely impractical for a realistic observing strategy, given such a large survey. This value is chosen to be similar to the average elevation in the MALS survey (Priv. comm.). For fields where such an elevation was not possible, the maximum elevation is assumed. }} of~$60\,{\rm deg}$} assuming a nominal observing time. To estimate the true on-source time per pointing, we split the time equally between pointings, and assume 25\% overheads (as for MeerKAT\footnote{\url{https://skaafrica.atlassian.net/wiki/spaces/ESDKB/pages/2457960584/2025+Call+for+Proposals}}). This results in an on-source time of 0.5h per pointing using which we scale the rms from the nominal observation time by combining in quadrature the confusion noise per pointing with the thermal noise (scaled $\propto 1/\sqrt{t}$, where $t$ is time).

\clhnew{From this, we have} an estimated per-pointing sensitivity. In reality, however, the rms in a mosaicked survey will be more complex. Factors such as (i) observational challenges in reducing data around bright sources; (ii) the convolution of images to a common resolution \citep[see e.g.][]{Hale2021}; (iii) flux scale offsets and (iv) the mosaicking of pointings will affect the measured rms within any survey. We attempt to mimic these through:
\begin{enumerate}
\item We artificially increase the rms around A-team sources~\citep{2020PASA...37...18W} assuming a 50\% peak increase in rms, tapering as a Gaussian of $\sigma=$ 2\deg \ to mimic the challenges of data reduction around such sources which can lead to large variations in rms close to such sources \citep[see e.g.][]{Hale2021, Shimwell2022}. 
\item Whilst a tapering resolution has been adopted in some surveys \citep[e.g.][]{Mauch2003, Intema2017, Duchesne2025}, others convolve images to a common resolution \citep[e.g.][]{condon1998nrao, Hale2021, Shimwell2022} prior to mosaicking to avoid smearing sources as pointings are combined. Through convolving to a common resolution, the angular resolution of the images is degraded, resulting in changes in the rms of the image. To provide conservative source estimates, we adopt the latter approach and estimate the common resolution rms through interpolating the relationship between resolution and sensitivity for different Briggs' weightings at the poorest resolution across the pointings.
\item Whilst each pointing will be calibrated, residual flux-scale offsets may remain within the image. The SKAO specifications\footnote{see \url{https://www.skao.int/en/science-users/118/ska-telescope-specifications}} require these calibration offsets to be limited to 1\% and so we \svhta{introduce} flux offsets per pointing to the rms (and additionally source flux densities) using a normal distribution centred at the rms and with $\sigma=$1\%.
\item Mosaicking of neighbouring pointings will improve sensitivity. We simulate such effects by assuming that all reduced images per pointing are truncated at the 30\% beam power (as used in e.g. LoTSS). We adopt a Gaussian modelling for sensitivity tapering from the central rms, using a model based on the FWHM of the primary beam, increasing the rms towards the edge of each pointing. For each simulated position across the field of view, the rms from all neighbouring pointings that position is observable from and weighted together.
\end{enumerate}

\begin{figure}
    \centering
    \includegraphics[width=0.7\linewidth]{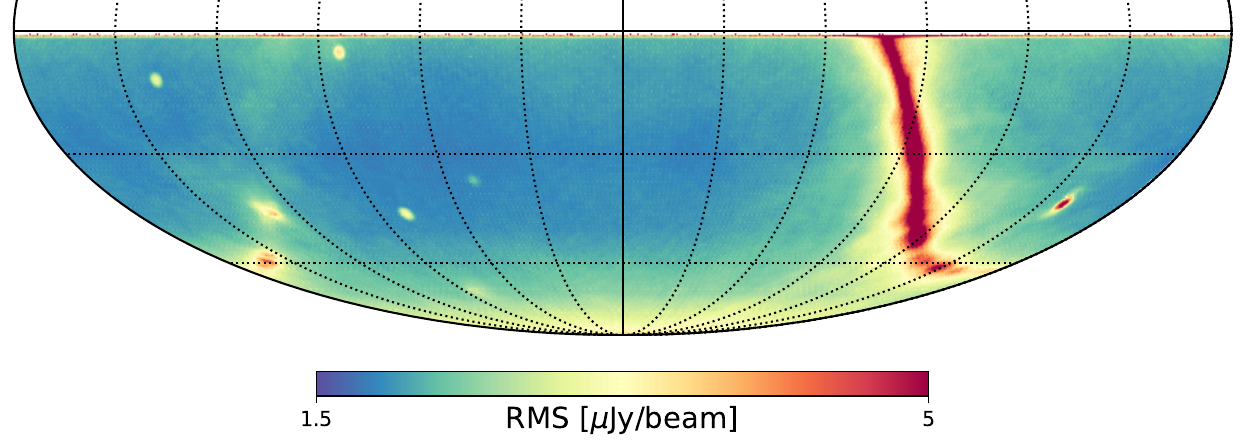}
    \caption{{Estimated rms distribution of a planned 20 000 sq deg, 10 000 h AA4 survey (incl. galaxy).}}
    \label{fig:rmsAA4}
\end{figure}

\cht{The varying sensitivity across the proposed survey area can be seen in Figure \ref{fig:rmsAA4}. We also generate a similar map for pointings which avoid the Galactic plane and go to higher declinations. The rms can be seen to increase towards the galactic centre and the southern pole, as well as near the A-team sources around which we have artificially increased the rms to reflect likely imaging challenges. }

\subsection{Simulation of \cht{Input} sources}
\label{sec:simulations}

\svht{To understand the expectation of such a survey, we generate mock catalogues and which, utilizing the rms map from the previous section, account for expected systematics in an SKAO like survey.}

\cht{Firstly, we generate a selection of potentially detectable input sources across the potential field of view for such a survey with the SKAO.} \svht{We sample source positions across the field of view from cosmological matter density fields simulated with \texttt{GLASS}~\citep{GLASS2023}, a simulation framework that efficiently generates large-scale structure realizations within the light cone from angular power spectra nested in user-defined radial shells (without the need for full N-body solvers). We fix the input cosmology to that of Planck 2018~\citep{Planck:2018vyg} and pre-compute angular power spectra with the help of \texttt{CCL}\footnote{\href{https://github.com/LSSTDESC/CCL}{https://github.com/LSSTDESC/CCL}}~\citep[Core Cosmology Library; ][]{LSSTDarkEnergyScience:2018yem} and \texttt{camb}~\citep[][]{Lewis:1999bs} in shells defined as top hat windows in the redshift range $z\in[0,4.5]$ and with equal spacing in comoving distance of \svhta{$\delta x=100\,{\rm Mpc}$}\footnote{Corresponding to redshift spacings ranging from $\delta z\approx0.05$ at low redshift to $\delta z\approx 0.33$ at the highest.}.  The angular resolution is further set by specifying $\ell_{\rm max}=3N_{\rm Side}-1$ with the \texttt{HEALPix}\footnote{\href{https://healpix.sourceforge.io}{https://healpix.sourceforge.io}}~\citep{Gorski:2004by} resolution $N_{\rm Side}=512$.}

\svht{We simulate the contribution of two sub-populations: active galactic nuclei (AGN) and star forming galaxies (SFGs) separately (see Section \ref{sec:radiobias}), owing to them being known to have differences in their luminosity functions and galaxy bias \citep[see e.g.][]{Hale2018, Best2023, Hale:2025wxp}.  The redshift distributions adopted are Gaussian kernel density estimates of T-RECS
} simulated galaxies from~\citet{Bonaldi:2023fnb}\footnote{Using the file \href{https://www.dropbox.com/scl/fo/fg83nwo03inoglgskr78m/AJik69sGh71o3i_mNvd6AQc?rlkey=b6ztqopwei6sj7nlpt8dmk2je&e=1&st=uxbauuko&dl=0}{\texttt{Mainrun\_cont/catalogue\_continuum\_clustered\_wrapped.fits}}}, \svhta{whilst for the galaxy bias $b_g(z)$ we adopt the functional form $b_{g}(z)=b^*_{g}/D(z)$, where $D(z)$ is the cosmic growth factor.  We fix $b^*_g=1.60$ for the AGN and $b^*_g=1.00$ for the SFG}, resulting from performing an error-weighted average of the biases found by~\citet{Hale:2025wxp} in the four lowest redshift bins ($b_{CC}$ in their Table~3).

\svht{Currently, \texttt{GLASS} is not able to produce source samples that respect a luminosity function with given functional form.  \svhta{As a work-around, we perform correlated draws from the T-RECS source property distribution as follows.}  Firstly, we adopt a nominal initial flux density limit of $S_{780\,{\rm MHz}}>1\,\mu{\rm Jy}$, roughly matching our $\mathcal{O}(\mu{\rm Jy})$ rms estimates --- this allows us to include the effects of faint sources which may be boosted into the catalogue through being co-located with a noise spike.  \svhta{We determine the expected source densities for AGN and SFGs separately, by extrapolating their numbers found in the $25\,{\rm deg}^2$ T-RECS footprint to the full sky, resulting in average number densities of $\overline{n}_{\rm AGN}=1.656\,{\rm arcmin}^{-2}$ and $\overline{n}_{\rm SFG}=34.04\,{\rm arcmin}^{-2}$.  This, combined with the above-mentioned choices in \texttt{GLASS}, results in the absolute numbers of sources $N_{\rm AGN}$ and $N_{\rm SFG}$ per simulation, with source positions, RA and Dec, in assigned redshift shells.  Our aim is now to assign to each source a set of properties that match its redshift shell.  To this effect, we begin to construct our sample by building a four-dimensional kernel density estimate\footnote{We use the \texttt{sklearn} code \texttt{KernelDensity} with a Gaussian kernel and set the bandwidth to bw=0.05.} (KDE) from source flux density $\log_{10}(S_{780\,{\rm MHz}})$, source size\footnote{We also exclude those T-RECS sources from the following that have source sizes $s=0$ (AGN only).} $\log_{10}(s)$, true redshift $\log_{10}(z)$, and spectral index\footnote{Interpolating the flux of neighbouring frequencies ($610$-$1400\,{\rm MHz}$) and obtain a per source spectral index $\alpha_{780\,{\rm MHz}}$.} $\alpha_{780\,{\rm MHz}}$, for AGN and SFG respectively.  From these we draw samples of sizes $N_{\rm AGN}$ and $N_{\rm SFG}$ which represent the same luminosity function, moreover the same source property distribution, as in T-RECS.  We match these properties to our \texttt{GLASS} sources in the correct redshift-ordering, by assigning their redshifts into \texttt{GLASS} redshift bins with a left-to-right sweep algorithm\footnote{This provides a greedy solution to the one-dimensional optimal transport problem on a path graph.}.  For purposes of computational efficiency in part \ref{sec:obseffects} of our simulation, we also assign to each \texttt{GLASS} source a T-RECS source ID, by querying a nearest-neighbour tree\footnote{We use \texttt{scipy.spatial.cKDTree}.}, built using the T-RECS sample.}}

\subsection{\cht{Simulating the Observational Effects into Output Catalogues}}
\label{sec:obseffects}

\cht{To simulate source detection (including realistic observational systematics) taking these steps:}
\cht{\begin{enumerate}
\item We obtain the weighted rms at the position of each simulated source (point 4 of Section \ref{sec:rmsmap}).
\item Using the combination of flux and shape properties\footnote{We note that as discussed in \cite{2021MNRAS.506.4121A}, the sizes for the largest AGN in T-RECS appears to large - this affects SNR and thus the detection probability for a source. Therefore for large sources, we also considered resampling so the size is more comparable to the major axes in the LoTSS-DR2. This should have limited effects in this survey though as the sources are dominated by SFGs.} for each simulated source we assign a simulated peak flux per source. We do this through simulating the source emission using the T-RECS prescription and adapted code from the \texttt{Simuclass} pipeline\footnote{\url{https://github.com/itrharrison/simuclass-public/}} and convolve the emission with a 2D Gaussian beam model, assuming a circular beam at the common angular resolution (point (2) of Section \ref{sec:rmsmap}) using a nominal pixel scale 1/5$^{\textrm{th}}$ of the PSF size.
\item Next, we apply a per-pointing flux-scale offset (see point (3) of Section \ref{sec:rmsmap}) using the randomly generated flux offset per pointing to obtain the input peak and integrated flux densities and obtain an input peak signal-to-noise ratio (SNR) per source using the rms.
\item We then utilise the simulations of \cite{Shimwell2022} as adopted in \cite{Hale2024} to generate the completeness probability for each source to be detected, given the input SNR. This accounts for the fact that source finding is not perfect and most source finders are just 50\% complete at 5$\sigma$ (the typical threshold of many radio surveys) as discussed in e.g. \cite{Boyce2023, Hale2024}\footnote{The simulations of \cite{Shimwell2022} use \texttt{PyBDSF} \cite{PyBDSF} as their source finder of choice.}. Following \cite{Hale2024} we randomly sample the completeness probability to determine whether a source would be detected or not.
\item \todo{For those sources considered `detected'}, we again use the simulations of \cite{Shimwell2022} and broadly follow the methodology of \cite{Hale2024} to obtain ``measured'' integrated and peak flux density measurements. This samples the distribution of SNR vs. input-to-measured flux densities as a function of input SNR\footnote{We note in \cite{Hale:2025wxp} this methodology was modified to obtain measured and integrated flux densities together, given they are correlated. However we adopted the existing outputs from \cite{Shimwell2022} compared to simulations over smaller fields of \cite{Hale:2025wxp} as they may contain additional systematics of large area surveys.}.  These ``measured'' flux values account for the source finder measurement biases due to factors such as the underlying noise causing flux boosting (or depleting). Using these measured flux densities, we can apply SNR and flux density cuts to the simulated survey sources which better mimic the result of additional cuts on a realistic sky survey as well as the source measurements themselves.
\item \svhta{We generate photometric redshifts $z_p$ from the true redshifts, $z$, by sampling from a Gaussian $z_p\hookleftarrow\mathcal{G}\big(z,\sigma_z(1+z)\big)$ following \cite{Harrison:2016stv}.}
\end{enumerate}
}

\subsection{Limitations in these simulations}

Whilst we have already outlined some challenges associated with radio cosmology studies in Section \ref{sec:challenges}, we note that whilst we have tried to understand the expectations in an SKAO-like survey, some challenges and limitations will remain - we discuss a few of them here. \dpt{For example, these simulations have assumed a uniformity and exact precision of mosaicking to a common resolution however in practice, larger flux calibration offsets and source smearing may affect sources which we have not modelled for in this case \citep[see e.g.][]{Shimwell2022, Shimwell2026}. This will vary both on the quality of the raw data itself and the data reduction processes. Therefore with real data we may be more subject to observational biases than we are able to model in our simulations, which could lead to some large-scale power systematic excess, unless it was corrected for or otherwise mitigated in some fashion.} \dpt{Moreover, these simulations have also assumed that the catalogue contains the correct identification of a single radio galaxy from multiple sources. As discussed in Section \ref{sec:challenges} \citep[and for a cosmology case in][]{2025PASA...42...62T}, the over- or under-association of sources by different source detection algorithms can lead to a systematic in the angular power spectra on small scales.} \svht{Finally, we did not consider model variations to the non-linear clustering assumed here~\citep{Mead:2020vgs}, and note that other models of non-linear growth of density fluctuations and even galaxy bias $b(z)\rightarrow b(k,z)$ might condition the source selection on small scales.  We leave this for future work.}

\section{Continuum Cosmology with the SKAO}
\label{sec:properties}

\begin{wrapfigure}{c}{0.45\textwidth}
    \centering
    \vspace{-0.4cm}
    \includegraphics[width=7cm]{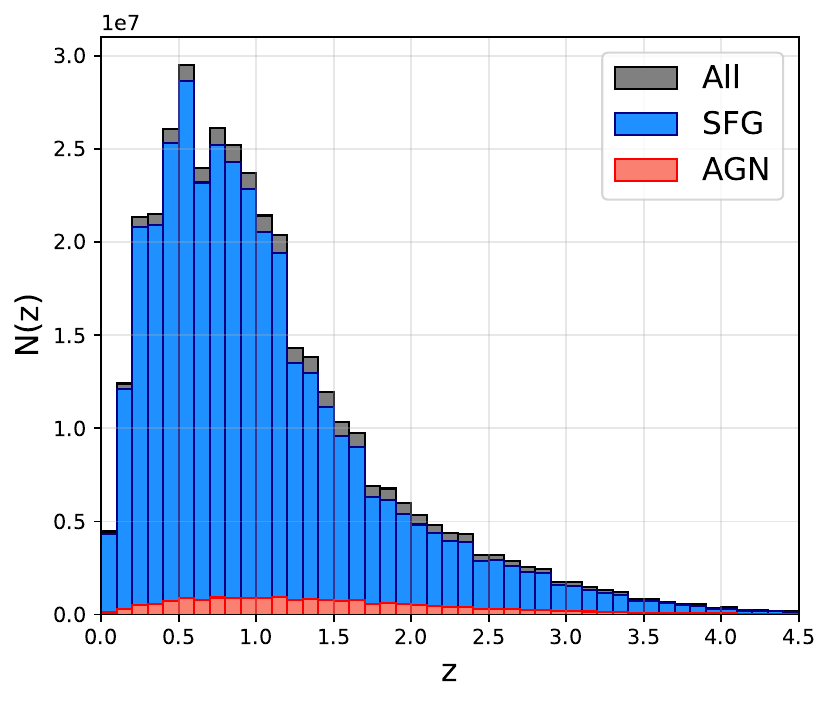}
    \caption{\clhnew{Redshift distribution ($N(z)$) of the expected radio sources (grey) and split into SFGs (blue) and AGN (red).}}
     \vspace{-0.5cm}
    \label{fig:pz}
\end{wrapfigure}

\cht{As a result of such simulations, a total of $\sim$\clhnew{380} million sources are expected to be detected in a similar style survey to that of the SKAO Red Book, covering 20 000 sq deg. Of these, $\sim$95 \% are SFG and the remaining sources as AGN. This fraction of SFG will decrease as increasingly higher flux cuts are applied \clhnew{\citep[see e.g.][]{2021MNRAS.506.4121A, Best2023}.} We present a comparison of the expectations on source numbers from such a survey in Figure \ref{fig:surveyscomparison} and estimations for the redshift distributions for such sources in Figure \ref{fig:pz}. In the left hand panel of Figure \ref{fig:surveyscomparison} we compare the expected rms sensitivity and area of this survey to both current wide area and deeper surveys (over $>$1 sq. deg) across telescopes such as ASKAP, LOFAR, MeerKAT and the VLA. This highlights that the planned survey would reach  depths comparable to that of some of the deepest surveys which exist to date, but over areas comparable to the widest area surveys to date and at higher resolution than current surveys from MeerKAT and ASKAP \citep[see e.g.][]{2025PASA...42...71H, Hale2025MIGHTEE} - thus overcoming the effects of confusion which affects deep surveys such as MIGHTEE \citep{Hale2025MIGHTEE}. Additionally, in the right panel of Figure \ref{fig:surveyscomparison} we present the cumulative source counts across the proposed survey, adding on the number density of existing surveys \citep[][]{condon1998nrao, Shimwell2022, Duchesne2025} to expectations of source numbers from the final EMU survey \citep{2025PASA...42...71H}. This highlights the increase in sources by a factor of 10 compared to the EMU survey \clhnew{and will represent} the largest number of radio galaxies that will be detected in the southern sky. This will be complemented in the north with low frequency surveys from LOFAR and the upgraded LOFAR2.0 telescope. }



\begin{figure}
    \centering
    \includegraphics[width=0.48\linewidth]{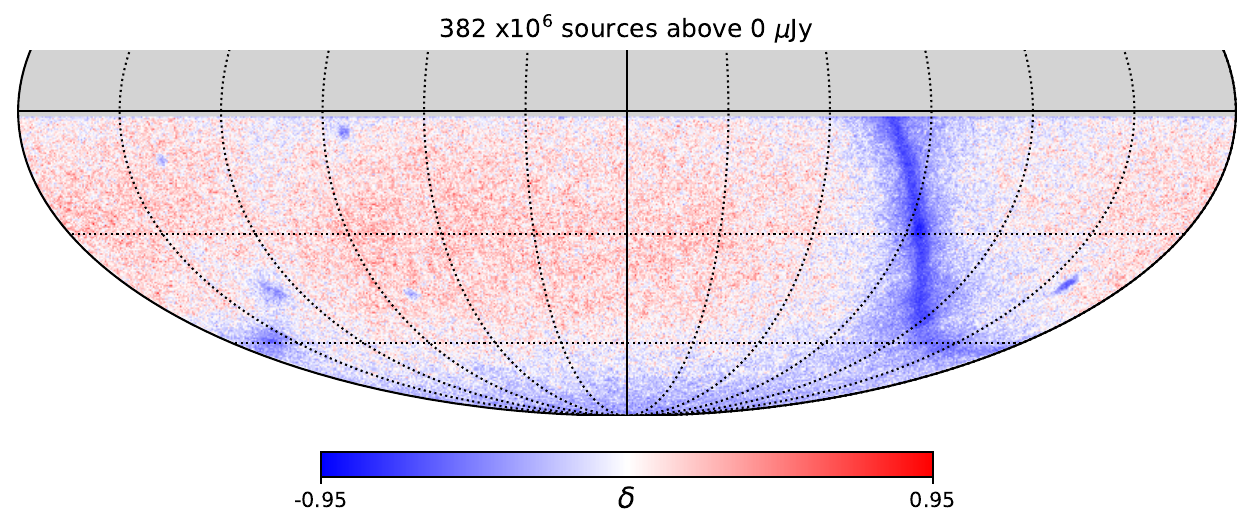}
    \includegraphics[width=0.48\linewidth]{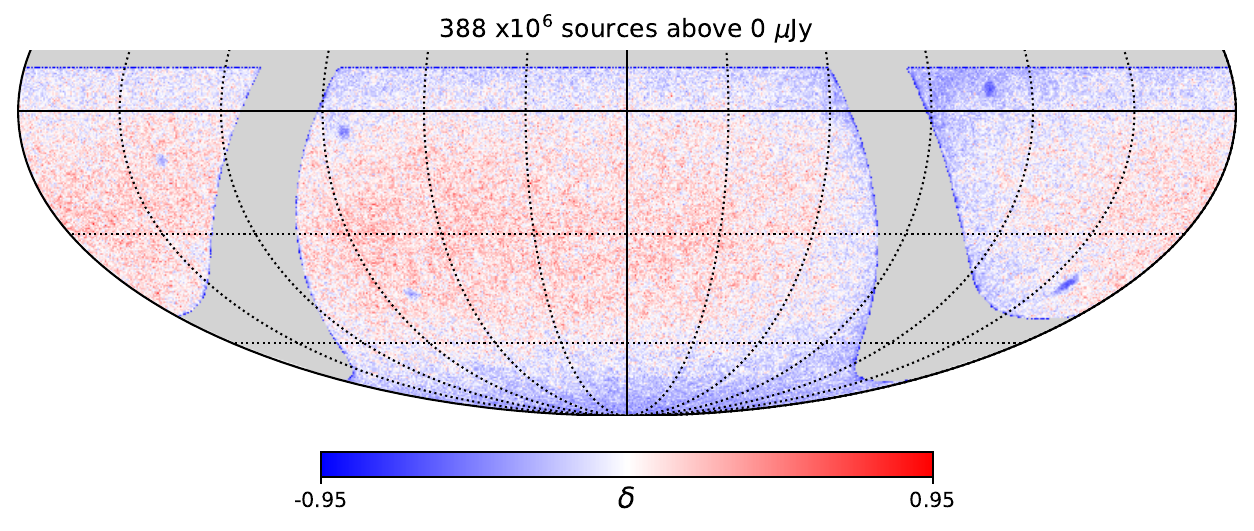}
    \newline
    \includegraphics[width=0.48\linewidth]{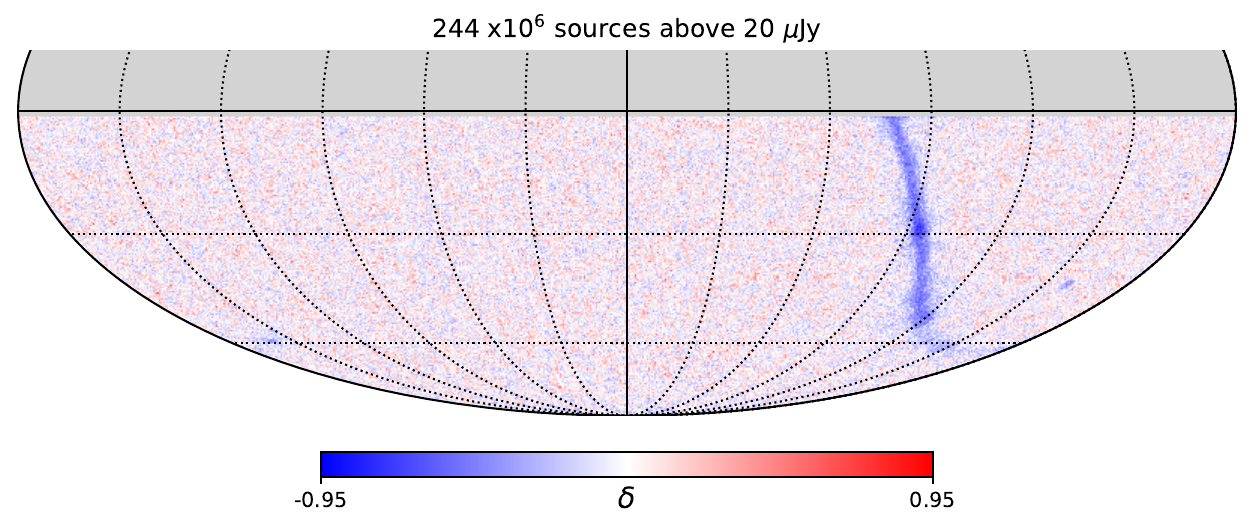}
    \includegraphics[width=0.48\linewidth]{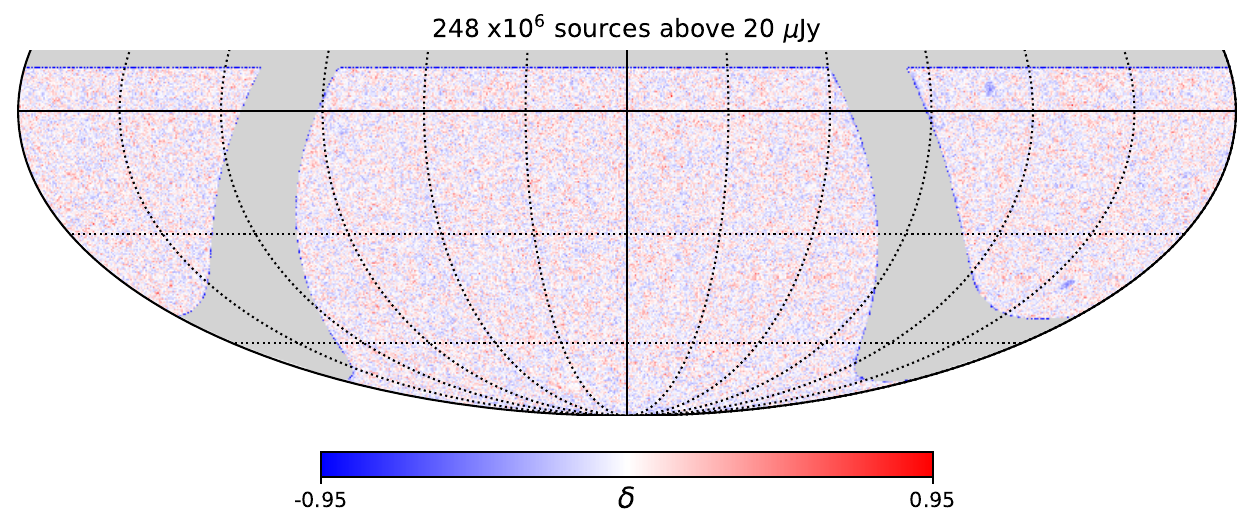}
    \newline
    \includegraphics[width=0.48\linewidth]{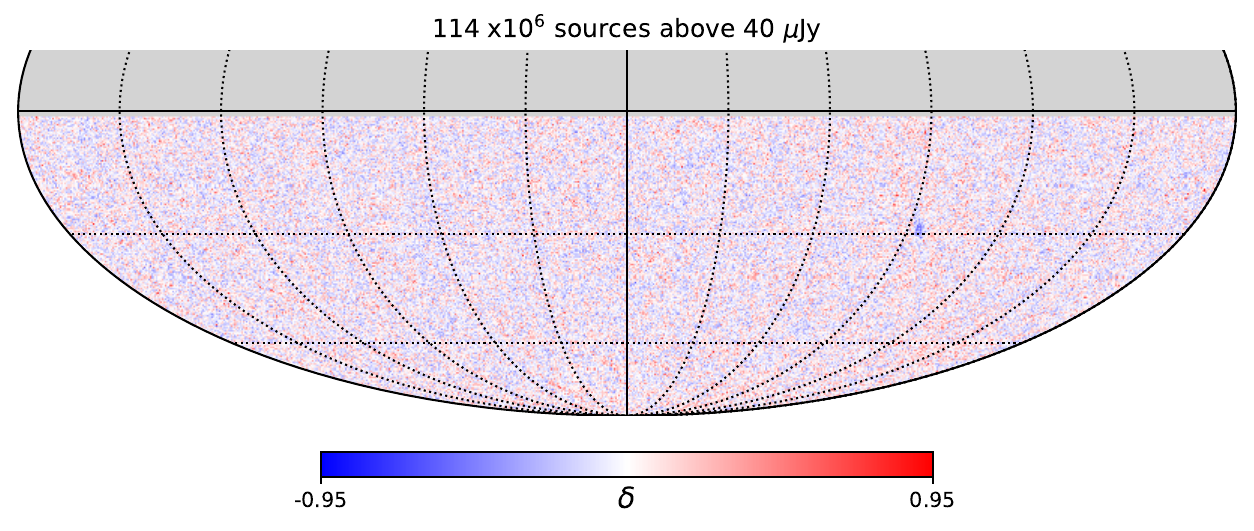}
    \includegraphics[width=0.48\linewidth]{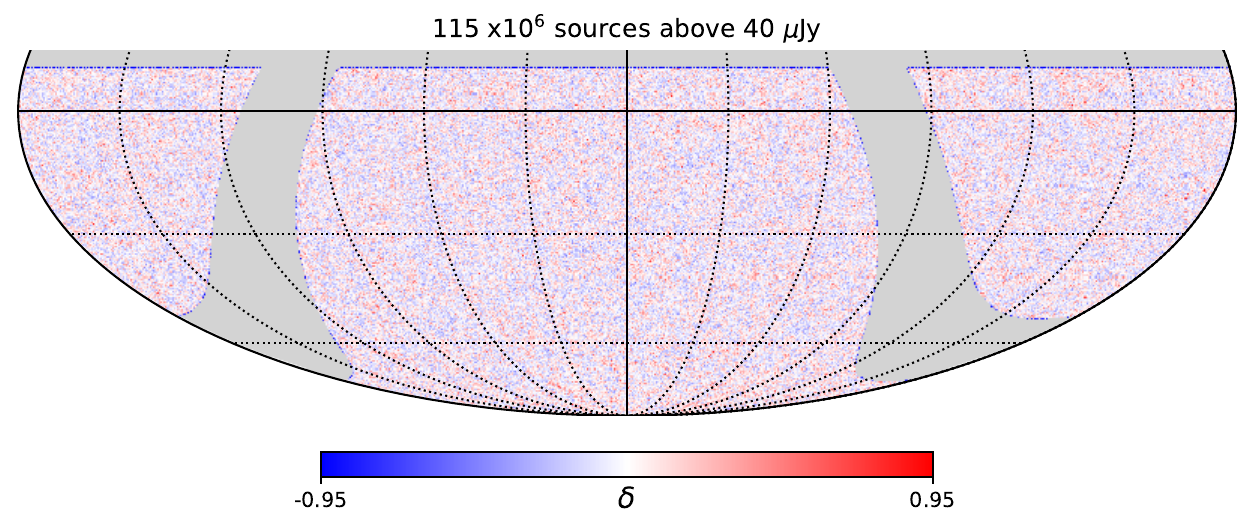}
    \caption{Comparison of the over-density ($\delta$) as in Equation \protect \ref{eq:density} for a 20 000 deg$^2$ survey which includes (left) or excludes (right) the Galactic plane with increasing flux density for no flux cut (top), 20$\mu$Jy cut (middle) and 40 $\mu$Jy (bottom). Included in the title of each plot is the number of sources expected in the survey.}
    \label{fig:density}
\end{figure}

\subsection{Homogeneity and Comparisons with the Red Book}
\label{sec:homogeneity}
Whilst $\sim$\clhnew{380} million sources are considered detected in this survey, these will not be homogeneously detected across the field of view, owing to the non-uniformity of the rms across the sky. We present this non-uniformity in terms of the density (see Equation \ref{eq:density}) in Figure \ref{fig:density}. This shows comparisons for when no flux limit, a 20$\mu$Jy flux limit and 40 $\mu$Jy flux limit is applied for both a 20 000 sq. deg survey including and excluding the Galactic plane. As can be seen, there are clear variations in source detections for the raw detected catalogue both near the Galactic plane and South Pole, this could lead to systematic effects in galaxy clustering if not accounted for (with weight maps) for cosmological analysis. \clhnew{Such variations (excluding over the Galactic plane) appear to be significantly reduced when a 20$\mu$Jy flux limit is applied. However when a 40$\mu$Jy flux density cut is applied the variations across the field reduce significantly, including over the Galactic plane}. As such, we suggest the ideal focus for a cosmology large area survey would avoid the Galactic plane and apply at least a 20$\mu$Jy limit. However, commensality with other SWGs may lead to the Galactic plane also being observed, in which case we suggest masking the Galactic plane or more conservative flux limits. \jab{We further validate this suggestion in section \ref{sec:Cls_SKAO} using $C_{\ell}$ measurements.}  

We note that in the Red Book, a 22.8$\mu$Jy limit was assumed, of which a prediction of $\lesssim$100 million sources was prediction. Instead with the 20$\mu$Jy limit, our surveys predicts $\sim$240 million galaxies (a factor of $\sim$2.5 more sources) which, in part, relates to the under-prediction of faint sources in SKADS as outlined in Section \ref{sec:intro}. This provides a significant increase in the number of sources available for continuum cosmology, though by a population whose galaxy bias is typically weaker than that of the brighter AGN \citep[see e.g.][]{Hale2018, Hale:2025wxp}.

\subsection{$C_{\ell}$ predictions for the SKAO}
\label{sec:Cls_SKAO}
\begin{figure}
    \centering
    \includegraphics[height=5.cm]{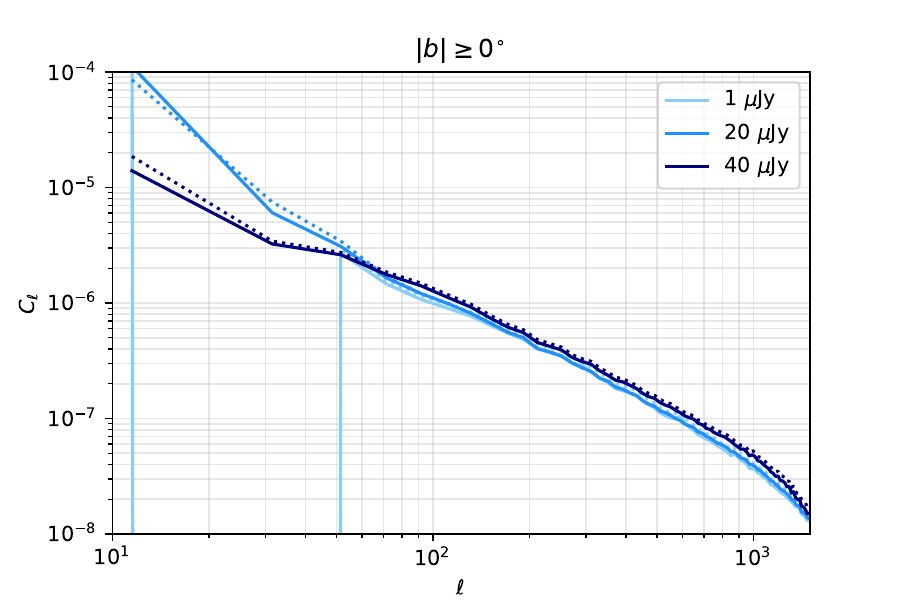}
    \includegraphics[height=5.cm]{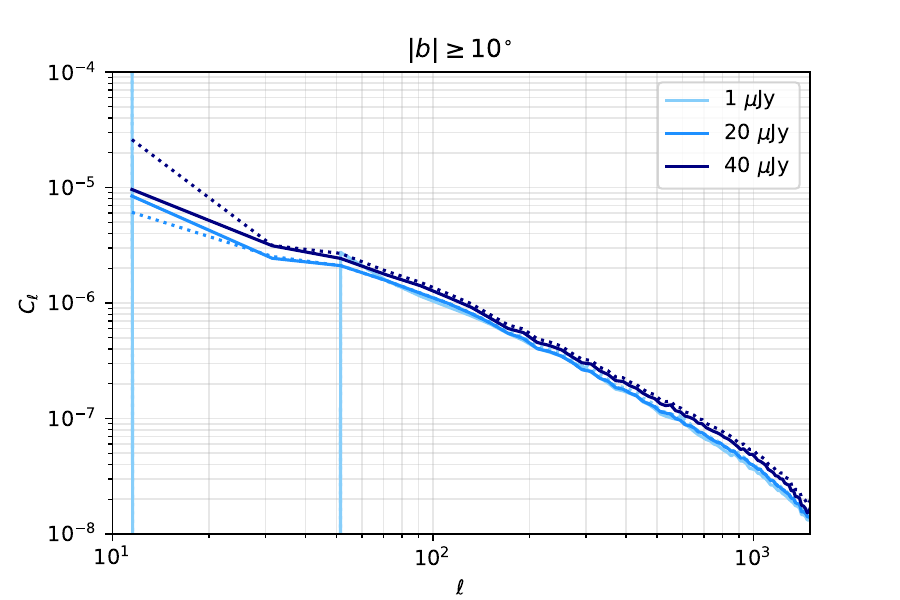}
    \includegraphics[height=5.cm]{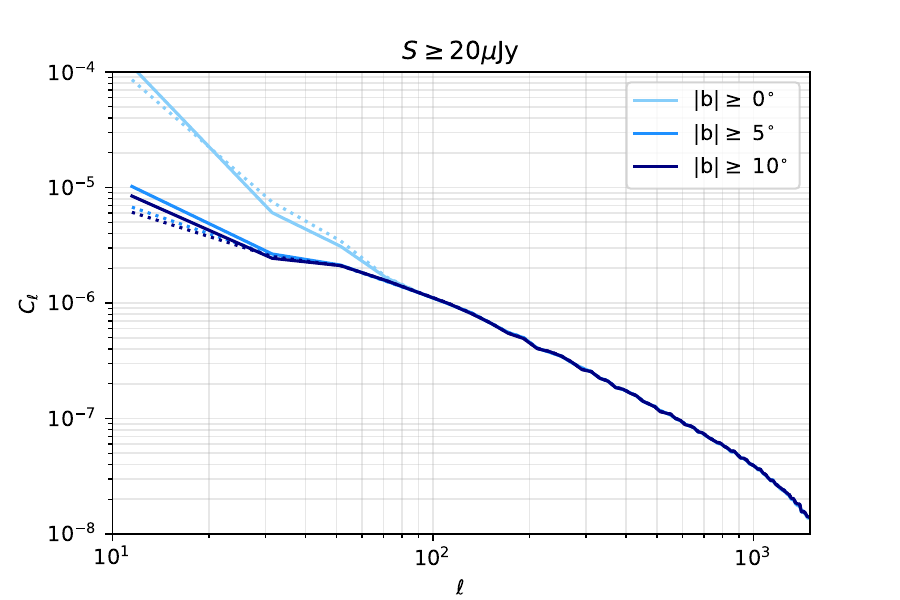}
    \caption{\clhnew{Comparison of the $C_{\ell}$ for a variety of flux density ($S$) and galactic latitude ($b$) cuts. Solid lines indicate $C_{\ell}$ measured from the output catalogue (solid lines) and the dotted lines indicate the $C_{\ell}$ measured using weights which account for incompleteness. Shown are: \textit{top left:} comparison when no galactic cut is applied; \textit{top right:} when a $\pm 10^{\circ}$ galactic cut is applied}; and \textit{bottom:} results for flux densities above $20 \mu$Jy using a variety of galacitc latitude cuts.}
    \label{fig:Cl_ch}
\end{figure}

\jat{In this section we present predictions for the large-scale structure measurements from the proposed survey with SKAO AA4 configuration. For that, we have measured the angular power spectra for the different simulated maps described in Section \ref{sec:sims}. As outlined above, the galaxy has strong synchrotron emission (see Figure \ref{fig:rmsAA4}) this induces fluctuation in the survey and spurious clustering. We demonstrate the effect that this has on the measured $C_{\ell}$ in Figure \ref{fig:Cl_ch} in order to validate our statements on the requirements for homogeneity in Section \ref{sec:homogeneity}. \jabnew{Using the simulations described in \ref{sec:simulations}, we create masks with weights given by the recovered random catalogue for a given flux cut with respect to the input random catalog, at each pixel. These masks allow us to correct for incompleteness in some cases.} In Figure \ref{fig:Cl_ch} we show raw $C_{\ell}$ measurements and weighted $C_{\ell}$'s which account for incompleteness using the detection fraction of sources - and so should represent the expectations from the input sources above such a flux limit \clhnew{(these represent the idealised case where all systematics are known and can be accounted for)}. \jab{We do not show analytic Gaussian error bars because on the plot as they are expected to be small, especially for $\ell>20$ (we present a more qualitative comparison of these errors to an EMU like survey in Fig \ref{fig:EMU_SKAO}. However, with real observations, there will likely be additional systematic uncertainties which we cannot account for in these simulations which, alongside the effects of 1-halo clustering at large $\ell$, will affect uncertainties in $C_{\ell}$ and likely increase their size. } Therefore, agreement between the weighted and unweighted $C_{\ell}$ demonstrate where we expect homogeneity. In the top left panel we consider the raw and weighted $C_{\ell}$ when no galactic cut is applied to the data for a variety of flux cuts. This highlights that with the exception of the 40$\mu$Jy cut, large differences are seen in the $C_{\ell}$, emphasising the effect that the galaxy has on our measurements. Instead if a $\pm10^{\circ}$ galactic latitude cut is applied (top right panel) such differences in the $C_{\ell}$ for the raw vs. weighted measurements are only significant when no flux density cuts are applied. Finally, taking our proposed 20$\mu$Jy cut cut we demonstrate the effect of the galactic cut, highlighting that even a $\pm 5^{\circ}$ may be suitable for cosmological analysis, if corrections are not applied to the data. Of course, for a real survey if the accurate weights for detection across the field of view can be obtained then lower flux density limits could be applied, increasing the number of sources available for cosmological analysis \citep[e.g. as used in][]{2022MNRAS.517.3785B, Hale2024}. }

\subsection{Comparisons to EMU}
\begin{wrapfigure}{r}{0.45\textwidth}
    \centering
    \vspace{-1.cm}
    \includegraphics[width=\linewidth]{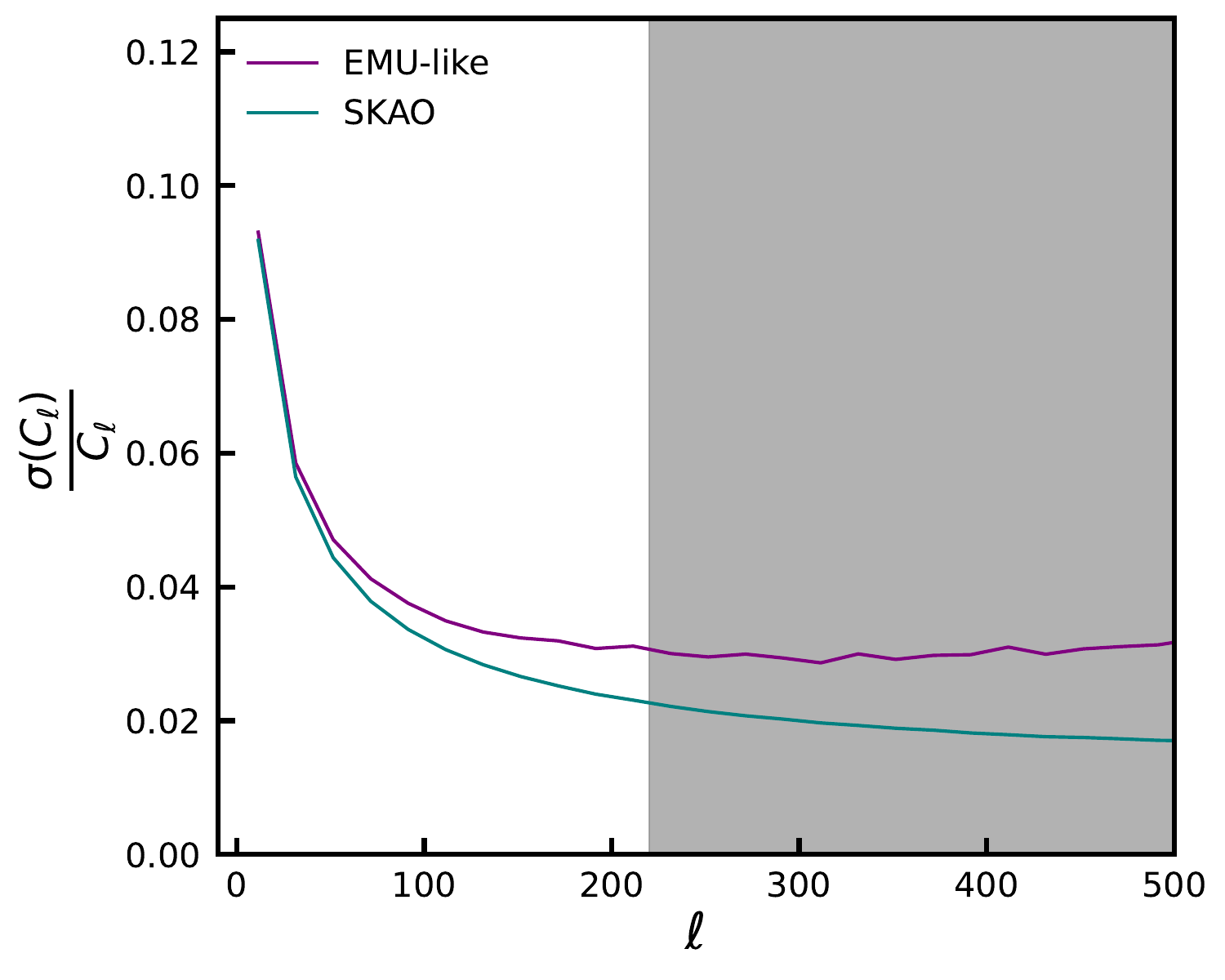}
    \caption{Relative error on the estimated angular power spectrum for the proposed SKAO AA4 continuum survey and an EMU-like survey. The grey area represents the non-linear scales.}
    \label{fig:EMU_SKAO}
    \vspace{-0.5cm}
\end{wrapfigure}

Until the proposed SKAO AA4 continuum survey is done, EMU will be the largest continuum survey in the southern hemisphere. In Figure \ref{fig:EMU_SKAO} we compare the relative error of the angular clustering, $\sigma(C_\ell)/C_\ell$, from an EMU-like survey will do with respect to our proposed SKAO AA4 survey. The errors are given by equation $\sigma(C_\ell)=[\text{Cov}(C_\ell,C_\ell)]^{1/2}$ where the covariance matrix of the angular power spectra $\text{Cov}(C_\ell,C_\ell)$ is defined in Equation \ref{eq:cov}. We see an improvement of order $\mathcal{O}$(1.5-2) with the proposed SKAO survey using pseudo-linear scales ($\ell<500$). We shade the non-linear area as our simulations do not include realistic non-linear evolution of galaxy clustering. Notice than here we are assuming the same beam resolution for both surveys while in reality SKAO will have a much better resolution ($2''$) than EMU ($15''$). We refer to \cite{Harrison01.2026.SKA} for the forecast on cosmological parameters when combining multiple continuum probes (although the sample is not exactly the same as the one proposed here) and \cite{Bertacca01.2026.SKA} for dipole measurements with a continuum survey. 

\subsection{1-halo clustering in the SKAO Era}
\label{sec:hod}

Whilst the 1-halo clustering of galaxies \citep[see e.g.][]{Zheng2005} has been explored thoroughly in the optical and IR \citep[see e.g.][]{Zehavi2011, Hatfield2016}, this is relatively unexplored in the case of radio continuum sources \citep[but see e.g.][]{Petter2024, Hamlett2026} and we have not modelled such an effect in our simulations. However, with deep radio surveys with current precursors, these 1-halo clustering effects are being observed \citep[see e.g.][]{Hale:2025wxp, Hamlett2026} owing to the depth and resolution of these surveys. Of course, with the surveys proposed in this work, these studies can be extended to much larger areas and provide much tighter constraints on how radio sources occupy singular dark matter haloes - providing key information especially for the rarest sources and the faintest objects. Such studies require multi-wavelength imaging in order to assign host galaxies, associate radio emission and split the radio sources both tomographically and as a function of source properties (see Sections \ref{sec:challenges}). To maximise this overlap with large surveys such as Euclid \citep{Euclid}, DESI \citep{DESI} and LSST \citep{LSST} is crucial. However, deeper fields will be crucial to pushing such analyses to broader populations (with a range of luminosities, stellar masses etc) and over a broader range of redshifts. Whilst this is not the scope of this survey, deep surveys over those fields with a wealth of ancillary information which are planned to be surveyed for other science cases (e.g. \cite{Prandoni01.2026.SKA} in AASKAII ) will be crucial in pushing these efforts forward.

\section{Summary}
\label{sec:summary}

The AA4 configuration of SKA can be used to perform a radio continuum survey of the entire Southern sky, using $\sim15\, 000$ individual pointings to cover $\sim20\,000\,{\rm deg}^2$. We utilised the SKAO continuum sensitivity calculator to obtain properties of such a survey (rms and angular resolution), and generated realistic mock catalogues that include the effect of spatial rms variations and incompleteness effects of source detection and characterisation across the field of view.

We found that such a survey can easily reach a flux limit that returns a number density of sources that is $10$ times larger ($\sim$ 200 million) than previous large-scale surveys and $3$ times larger than predicted in the SKA Cosmology Red Book. By measuring the angular clustering power spectrum of these simulated sources in a single redshift bin for our proposed SKAO AA4 survey, we found that a survey strategy that avoids the Galactic plane by targeting regions with galactic latitude $|b|>5º$ can reach a flux limit of $20 \mu$Jy and expect source detection isotropy across the survey footprint. Finally, we compare the proposed SKAO AA4 survey with the EMU-like precursor and show that we can reach an improvement of $\sim 1.5-2$ in the relative error of clustering measurements within the linear scales, assuming the same angular resolution.


\section{Acknowledgments}
The authors want to thank S. Camera, I. Harrison, M. Spinelli for insightful discussions and T. Shimwell for providing LOFAR simulations. We also acknowledge the use of the computational resources of: Titan at CAPA Institute and CESAR at BIFI Institute (University of Zaragoza); Glamdring (University of Oxford); Cuillin (Royal Observatory, Edinburgh).

JA ackowledges the support of the grants PGC2022-126078NB-C21 and PID2024-160228NB-I00 funded by MCIN/AEI/10.13039/ 50110001103 and Diputación General de Aragón-Fondo Social Europeo (DGA-FSE) grant 2023-E21-23R funded by Gobierno de Arag\'on.
CLH acknowledges support from the Oxford Hintze Centre for Astrophysical Surveys which is funded through generous support from the Hintze Family Charitable Foundation and from the Science and Technology Facilities Council (STFC) through grant ST/Y000951/1.
SvH is supported by a Leverhulme Trust
national Professorship Grant to S. Sondhi (No. LIP-2020-014). 
BB-K acknowledges support from INAF for the project `Paving the way to radio cosmology in the SKA Observatory era: synergies between SKA pathfinders/precursors and the new generation of optical/near-infrared cosmological surveys' (CUP C54I19001050001).

\clhnew{The authors are ordered by contribution.}

\bibliographystyle{abbrvnat-maxbibnames4}
\bibliography{cosmo_continuum_auto,cosmo_swg_bib} 









\end{document}